\pdfoutput=1
\documentclass{PoS}

\usepackage{xspace}
\usepackage{subfig}
\usepackage{placeins}
\usepackage{siunitx}
\usepackage{float}
\usepackage{cite}
\usepackage{txfonts}
\usepackage{xcolor}
\usepackage{lineno}

\newcommand*{\POWHEG}{\textsc{Powheg}\xspace}
\newcommand*{\SHERPA}{\textsc{Sherpa}}
\newcommand*{\PYTHIA}{\textsc{Pythia}}
\newcommand*{\MADSPIN}{\textsc{MadSpin}}
\newcommand*{\MGMCatNLO}{\textsc{MadGraph5}\_aMC@NLO\xspace}
\newcommand*{\antibar}[1]{\ensuremath{#1\bar{#1}}\xspace}
\newcommand*{\ttbar}{\antibar{t}}
\newcommand*{\TeV}{\ifmmode {\mathrm{\ Te\kern -0.1em V}}\else
                   \textrm{Te\kern -0.1em V}\fi}%
\newcommand*{\GeV}{\ifmmode {\mathrm{\ Ge\kern -0.1em V}}\else
                   \textrm{Ge\kern -0.1em V}\fi}%
\newcommand*{\pt}{\ensuremath{p_{\rm T}}\xspace}
\newcommand*{\pT}{\ensuremath{p_{\rm T}}\xspace}
\newcommand{\stat}{\ensuremath{\:\textrm{(stat.)}}}
\newcommand{\syst}{\ensuremath{\:\textrm{(syst.)}}}

\newcommand{\theo}{\ensuremath{\:\textrm{(theo.)}}}
\newcommand*{\ifb}{\ensuremath{\textrm{fb}^{-1}}\xspace}
\newcommand*{\MET}{\ensuremath{E_{\mathrm{T}}^{\mathrm{miss}}}\xspace}
\renewcommand*{\to}{\ensuremath{\rightarrow}\xspace}
\newcommand{\coll}[1]{#1 Collaboration}
\newcommand{\arxiv}[2]{arXiv:\href{http://www.arxiv.org/abs/#1}{\color[rgb]{0.,0.7,0.}{#1 [hep-#2]}}}
\newcommand{\pub}[2]{\href{http://dx.doi.org/#2}{\color[rgb]{0.,0.7,0.}{#1}}}
\newcommand{\subm}[1]{submitted to #1}
\newcommand{\EPJC}{Eur.\ Phys.\ J.\ C}
\newcommand{\Mpole}{\ensuremath{m_t^{\text{pole}}}\xspace}
\newcommand{\Mrun}{\ensuremath{m_t (m_t)}\xspace}
\newcommand{\Rhos}{\ensuremath{\rho_{\text{s}}}\xspace}
\newcommand{\ttbarjet}{{\ensuremath{t\bar t + 1\textnormal{-jet}}\xspace}}
\def\MSbar{\ensuremath{\overline{\mbox{MS}}}\xspace}
\newcommand{\Robs}{\ensuremath{\mathcal{R}}\xspace}
\newcommand{\RobsTh}{\ensuremath{\mathcal{R}^{\ttbarjet}_{\text{NLO+PS}}}\xspace}

\title{Top-quark properties at ATLAS with focus on the latest mass and spin
correlation measurements}

\ShortTitle{Top-quark properties}

\author{\speaker{Markus Cristinziani}\thanks{Supported by the European Research
Council grant ERC--CoG--617185 and by the German Federal Ministry of Education and
Research (FSP-103)}\\
{\rm On behalf of the ATLAS Collaboration}\\
Physikalisches Institut, Universit\"at Bonn, Nussallee 12, 53115 Bonn, Germany.\\
E-mail: \email{cristinz@uni-bonn.de}}

\abstract{Properties of the top-quark are presented, with emphasis on the most
recent ATLAS measurements of the mass and \ttbar spin correlations, obtained
with proton--proton collision data collected at the Large Hadron Collider.
Normalised differential distributions are used in both cases.  For the
extraction of the top-quark mass, \ttbarjet\ single-lepton events are selected
from the 20.2~$\text{fb}^{-1}$ $8\,\TeV$ dataset, and the unfolded distribution at
parton level is compared with theoretical predictions to obtain
$m_t^{\text{pole}} =  171.1^{+1.2}_{-1.1} \GeV$ in the pole-mass scheme
and $m_t(m_t) = 162.9^{+2.4}_{-1.6} \GeV$ in the running-mass scheme.
For the measurement of spin correlations in \ttbar production, dilepton events
are selected using 36.1 fb$^{-1}$ $13\,\TeV$ data.  The azimuthal opening angle
between the two leptons is measured inclusively and as a function of the
invariant mass of the \ttbar system.  The observed degree of spin correlation
is significantly higher than predicted by the generators used, but agrees well
with the prediction of one of the fixed-order calculations.  }

\FullConference{7th Annual Conference on Large Hadron Collider Physics --- LHCP2019\\
		20--25 May, 2019\\
		Puebla, Mexico}

\begin{document}

\section{Introduction}

The mass of the top quark, the heaviest known elementary particle, is a key
parameter of the Standard Model (SM) of particle physics and must be determined
experimentally. In the SM, the gauge structure of the interaction of the top
quark with other particles establishes a relation between the top-quark,
Higgs-boson and $W$-boson masses. A precise determination of these three
parameters forms an important consistency check of the SM, and provides
information about the evolution of the Higgs quartic coupling, which affects
the shape of the Higgs potential and is associated with the stability of the
quantum vacuum.  Owing to the large mass, the lifetime of the top quark is
shorter than the timescale for hadronisation (${\sim}10^{-23}$~s) and is much
shorter than the spin decorrelation time (${\sim}10^{-21}$~s). As a result, the
spin information of the top quark is transferred directly to its decay
products.  Top-quark pair production (\ttbar) in QCD is parity invariant and
hence the top quarks are not expected to be polarised in the SM; however, the
spins of the top and the anti-top quarks are predicted to be correlated.  The
determination of \ttbar spin correlation is a sensitive test to physics beyond
the Standard Model (BSM), since the latter might modify the observed level of
correlation. 

\section{Measurement of the top-quark pole mass in \ttbarjet\ events}

A quantitative statement about the value of a quark mass requires a precise
reference to the mass scheme in which the mass is defined.  The mass scheme
which is used most often in top-quark mass measurements is the pole-mass
scheme, where the renormalised top-quark mass (the pole mass, \Mpole) coincides
with the pole of the top-quark propagator.  From the \ttbar production cross
section, on other hand, the running top-quark mass in the modified minimal
subtraction scheme (\MSbar) has been extracted. The two mass schemes can be
related precisely, with up to four-loop accuracy.
 
Direct top-quark mass measurements at hadron colliders, based on the
reconstruction of the top-quark decay products and using Monte Carlo (MC) event
generators to extract the mass, are frequently interpreted as the pole mass.
Recent works estimate that such an interpretation is affected by a $0.5-1 \GeV$
uncertainty due to non-perturbative effects.  With direct top-quark mass
measurements reaching sub-percent precision it becomes important to evaluate
uncertainties associated with the interpretation of the measured mass at the
same level of accuracy.  It is therefore important to extract the top-quark
mass by comparing data with predictions computed in a well-defined mass scheme.
In this case the ambiguity related to the top-quark mass interpretation is
avoided, allowing a precise evaluation of the uncertainty associated with the
mass scheme chosen.  In such measurements the MC event generator is only used
to correct distributions obtained from measured data for effects originating
from the detector and the modelling of non-perturbative physics.
 
ATLAS reported measurements of the top-quark mass in both, the pole-mass and
\MSbar schemes, taking advantage of the sensitivity to the top-quark mass of
the differential cross section of \ttbar production in association with at
least one energetic jet~\cite{polemass}. The measurement is performed using
$20.2\,\ifb$ of $8 \TeV$ $pp$ collisions collected in 2012.  The presence of
the additional jet enhances the sensitivity to the top-quark mass in comparison
with similar observables defined for the \ttbar system only.  In particular,
the observable used to extract the top-quark mass, \Robs, is defined as the
normalised differential \ttbarjet\ cross section as a function of the variable
$\Rhos = 2 m_0/m_{\ttbarjet}$, $\mathcal{R}(\Mpole, \Rhos) = (1 /
\sigma_{\ttbarjet}) \cdot \text{d}\sigma_{\ttbarjet~} / \text{d}\Rhos$, where
$m_0$ is a constant fixed to $170 \GeV$ and $m_{\ttbarjet}$ is the invariant
mass of the \ttbarjet\ system.  The normalised differential cross sections are
presented at the so-called particle level in which data are only unfolded for
detector effects and at the parton level where \Robs can be directly compared
with available fixed-order calculations.  The large $8 \TeV$ dataset makes it
possible to achieve a high precision in the measurement of the \Robs
distribution, in particular in the region where it is most sensitive to the
top-quark mass, ultimately allowing the top-quark mass to be extracted with
high accuracy.
 
The nominal $\ttbar$ sample that is used to unfold the data was generated using
the \POWHEG-hvq package, which is based on next-to-leading-order (NLO) QCD
matrix elements.  The event selection targets the single lepton \ttbar
signature with one reconstructed electron or muon and at least five jets, two
of which must be $b$-jets. After requirements on the transverse momentum (\pT),
the missing transverse momentum $\MET$ and the $W$-boson transverse mass, about
\num{28000} events are selected, of which 93\% are expected to be \ttbar events
according to simulation.

The \ttbarjet\ system is then reconstructed. Candidates for the hadronically
decaying $W$ boson are formed by pairing all jets not tagged as $b$-jets and
selecting pairs $i,j$ that satisfy $0.9<m_W/m_{ij} < 1.25$ and
$\text{min}\left(\pT^{i},\ \pT^{j}\right) \cdot \Delta R_{ij} <90\GeV$  where
$\pT^{i}$ is the transverse momentum of the jet $i$, $m_{ij}$ is the invariant
mass of the jet pair, $\Delta R_{ij}$ their angular distance.  The application
of these two requirements reduces the multijet and combinatorial backgrounds.
The neutrino momentum is reconstructed, up to a twofold ambiguity, by
identifying the \MET with its transverse momentum and using the $W$-boson mass
constraint to infer its longitudinal momentum.  Only events where at least one
neutrino candidate exists are considered.  If there are two solutions, each of
the neutrino candidates is added to the charged lepton, leading to two
$W$-boson candidates.

Pairs of hadronic and semileptonic top-quark candidates are formed by combining
all the hadronic and leptonic $W$-boson candidates with the two $b$-tagged
jets.  Among all possible combinations the one selected is that which minimises
the absolute difference between the masses of the reconstructed hadronic top
($m_{t_{\text{had}}}$) and the semileptonic top ($m_{t_{\text{lep}}}$)
candidates, divided by their sum, $\left|
m_{t_{\text{lep}}}-m_{t_{\text{had}}}\right| /
(m_{t_{\text{lep}}}+m_{t_{\text{had}}})$.  The \ttbar candidates must satisfy
$m_{t_{\text{lep}}}/m_{t_{\text{had}}}> 0.9$.  The four-momenta of the jets
which are identified with the hadronic decay of the $W$ boson are corrected by
the factor $m_W/m_{ij}$.  Among the jets not used in either top-quark
candidate, the leading jet in \pT is taken as the jet produced in association
with the top quarks, before their decay.  Only events where this extra jet has
a transverse momentum larger than $50 \GeV$ are considered. The purity of the
sample is more than $90\%$.  The \ttbarjet\ system is reconstructed adding the
four-vectors corresponding to the $b$-jets, the selected $W$-boson candidates
and the additional jet.  The inclusive quantity \Rhos is insensitive to
ambiguities in the combinatorics and is not affected by an incorrect pairing of
$b$-jets with $W$-boson candidates.  
 
The measured \Robs distribution is unfolded and compared with theoretical
predictions at fixed order, allowing the determination of the top-quark mass in
a well-defined theoretical framework.  In addition, the \Robs distribution is
also presented at particle level, where data are unfolded for detector effects
only.  This will allow direct comparisons with possible future theoretical
calculations which include top-quark decay and hadronisation effects.  The
unfolding procedure is detailed in the following.  First, the detector-level
distribution of \Rhos is re-binned to maximise the sensitivity of the
observable to the top-quark mass while keeping enough statistics in each bin.
This is achieved by choosing a fine binning in the region $\Rhos \gtrsim 0.6$,
where the observable is most sensitive to the top-quark mass.  Second, the
predicted background contribution is subtracted and the distribution is
normalised to unity.  Finally, the distribution is unfolded using iterative
Bayesian unfolding.
 
\begin{figure}[htbp]
\centering
\hfil
\subfloat[]{\includegraphics[height=0.245\textheight]{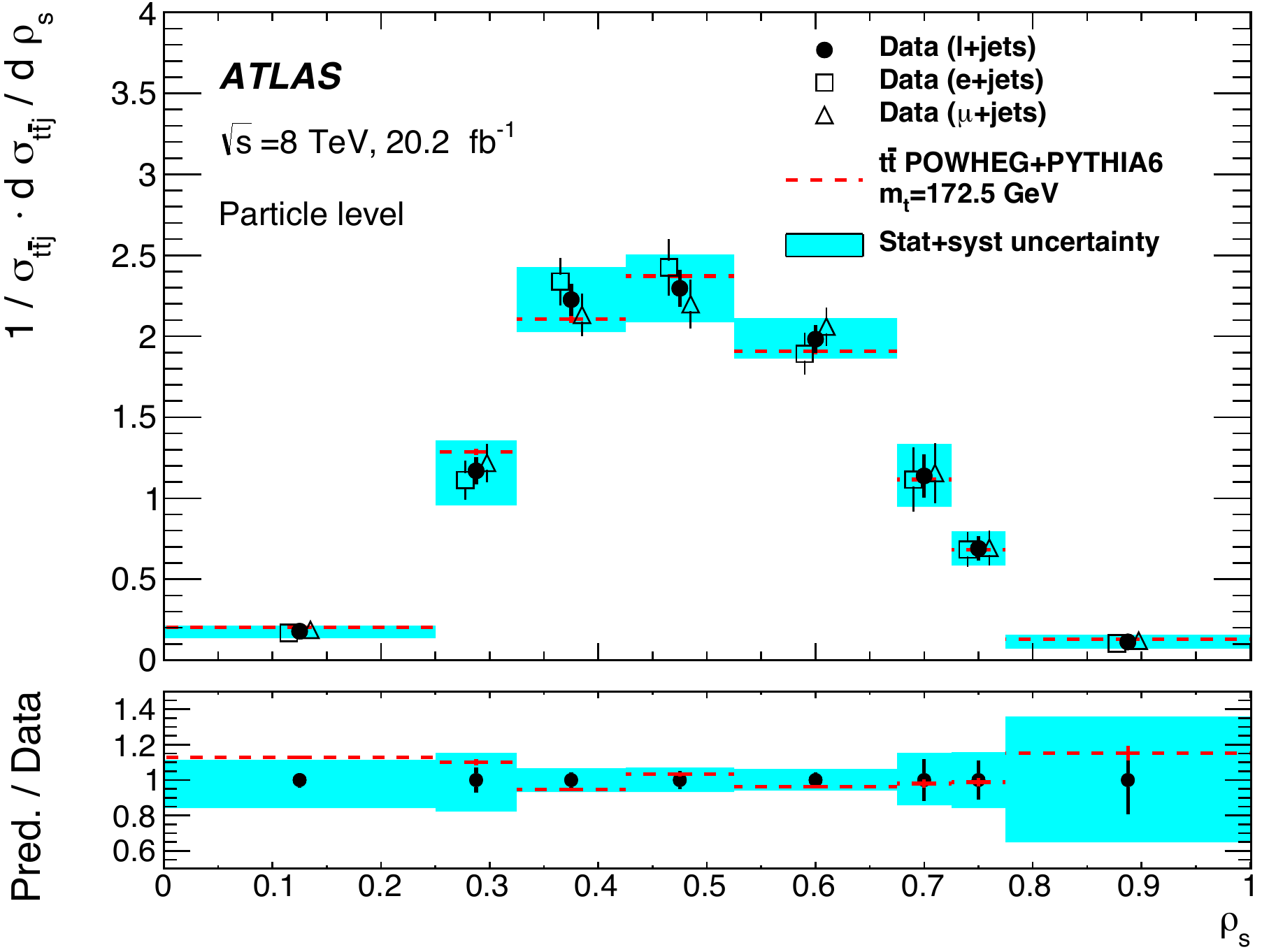}}
\hfill\hfill
\subfloat[]{\includegraphics[height=0.245\textheight]{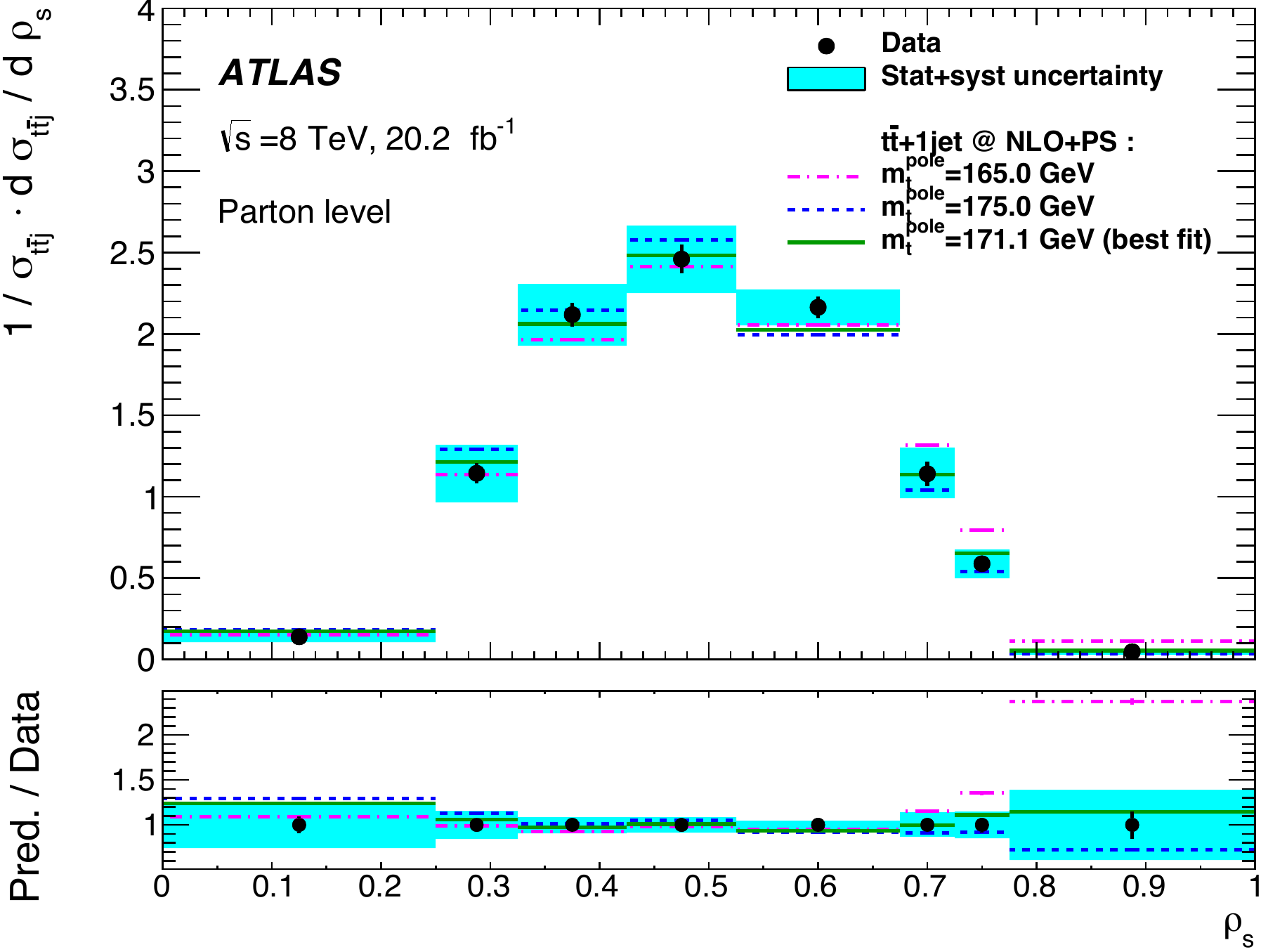}}
\hfill
\caption{%
The normalised differential cross section for $pp \rightarrow \ttbarjet$
production as a function of \Rhos~\cite{polemass}.  (a) The results in the
electron and muon channels, and the combination of the two, are shown.  The
data are unfolded to the particle level and are compared with the prediction
from \POWHEG + \PYTHIA 6.  (b) The data are unfolded to the parton level. The
predictions of the NLO+PS calculation are shown for various top-quark pole mass
assumptions.}
\label{fig:pole}
\end{figure}

In Figure~\ref{fig:pole} the unfolded, normalised differential cross section at
particle level is compared with the prediction of the \POWHEG + \PYTHIA 6
generator with the top-quark mass parameter set to 172.5~\GeV.  The
distributions obtained from the electron and muon channels separately, unfolded
following the nominal procedure, are also presented in the same figure to show
their compatibility with the combined result.  The same measurement is
presented after unfolding to parton level. The result is compared with the
prediction for \ttbarjet~production. The fixed-order calculation at NLO
accuracy in QCD is interfaced to the parton shower and is labelled as
``NLO+PS'' in the following.  The prediction is shown for two values of the
top-quark pole mass, to demonstrate the sensitivity of the observable to the
top-quark mass.

The top-quark pole mass is extracted from the parton-level result with an
NLO+PS calculation of $\ttbarjet~$ production.  In each bin of the distribution
a continuous parameterisation $\RobsTh (\Mpole)$ is obtained by interpolating
with a second-order polynomial between different \RobsTh predictions computed
at fixed \Mpole values.  The fit to the parton-level differential cross section
yields the top-quark pole mass $\Mpole = 171.1 \pm 0.4 \stat \pm 0.9 \syst
^{+0.7}_{-0.3} \theo \GeV$.  The result for the running mass in the \MSbar
scheme is $m_t(m_t) = 162.9 \pm 0.5 \stat \pm 1.0 \syst ^{+2.1}_{-1.2} \theo
\GeV$.  Several tests are performed to verify the consistency and robustness of
the result, including the choice of \pt{} cut on the additional jet and of the
$\Rhos$ binning.  The measured top-quark mass is independent of the assumed
top-quark mass in the MC simulation that is used to unfold the data, and is
found to be compatible in the electron and muon channels, separately.

\section{Comparison with other ATLAS top-quark mass measurements}

Figure~\ref{fig:summary}(a) shows a summary of measurements of the top-quark
pole mass.  The top-quark pole mass result obtained from data unfolded to
parton level is compatible with previous measurements.  Compared with the
result obtained by ATLAS with the same method at 7~\TeV{} the statistical and
systematic uncertainties of the new result are reduced by more than a factor of
two, yielding a relative uncertainty of 0.7\%, constituting the most precise
measurement of the top-quark pole mass with 8~\TeV{} data. With 13~\TeV{} data,
the CMS collaboration further improved the precision to 0.5\%, using
differential spectra in the \ttbar dilepton channel. The measurements of the
top-quark pole mass are competitive with the conventional methods, which are
summarised in Figure~\ref{fig:summary}(b). The most recent combination was
performed by the ATLAS collaboration~\cite{mass} using the 7 and 8~\TeV{}
\ttbar measurements, allowing to reduce the total uncertainty to 0.3\%,
equalling the measurement of the CMS collaboration that was released three
years before. With the analysis of the full Run-2 dataset it is to be expected
that both, the conventional methods as well as the extraction of the pole mass
will further be improved and attain new levels of precision.

\begin{figure}[htbp]
\centering
\subfloat[]{\includegraphics[width=0.49\textwidth]{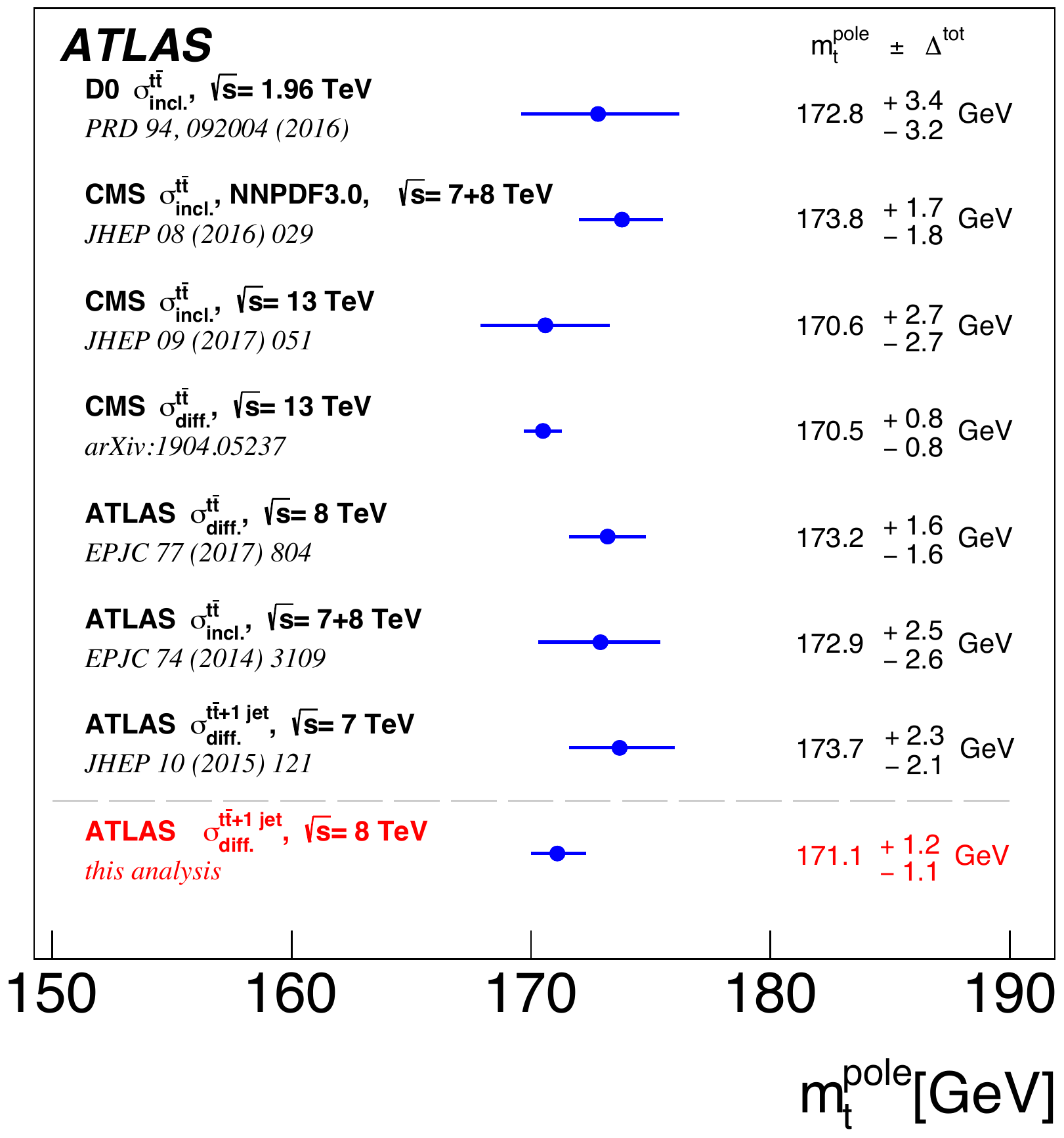}}
\hfill\hfill
\raisebox{2.7ex}{\subfloat[]{\includegraphics[width=0.47\textwidth]{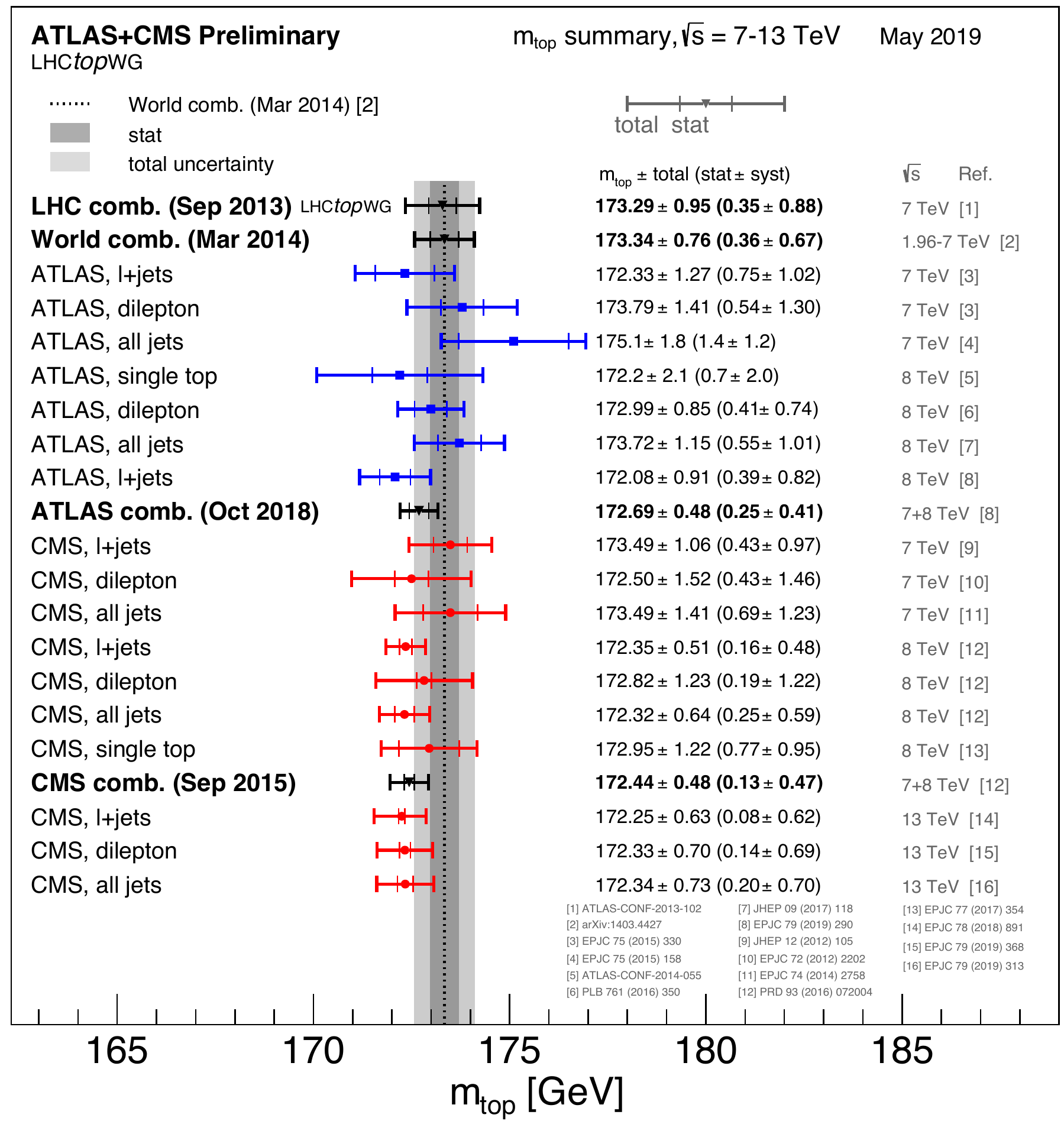}}}
\caption{%
(a) Summary of top-quark pole mass measurements at the Tevatron and the
LHC~\cite{polemass}. (b) Summary of ATLAS and CMS direct $m_t$ measurements.
The results are compared with the LHC and Tevatron+LHC $m_t$
combinations~\cite{properties}.} \label{fig:summary}
\end{figure}
 
\FloatBarrier

\section{Spin correlations} 

Spin correlations in the \ttbar system have been observed experimentally by the
ATLAS and CMS collaborations using Run-1 data. A new measurement has been
presented by ATLAS~\cite{spin}, using $13\TeV$ data collected in 2015 and 2016.
The spin information of top-quarks can be accessed through their decay
products.  Charged leptons arising from leptonically decaying $W$ bosons carry
almost the full spin information of the parent top quark and can be readily
identified and reconstructed.  Thus, observables to study spin correlation in
\ttbar events are often based on the angular distributions of the charged
leptons in the dilepton channel. Observables used in this measurement are the
absolute azimuthal opening angle measured in the transverse plane
($\Delta\phi$), and the absolute difference between the pseudorapidities
($\Delta\eta$).  Improved MC generators are employed relative to previous spin
correlation results to better control the systematic uncertainties. The spin
correlation is measured as a function of the invariant mass of the \ttbar
system, as well as inclusively.

The primary \ttbar MC sample used (\emph{nominal}) was simulated using the NLO
\POWHEG-Box matrix-element event generator interfaced to \PYTHIA\ for the
parton shower and fragmentation.  An alternative \ttbar sample was simulated
with the same settings but with the top quarks decayed using \MADSPIN\ and with
spin correlations between the $t$ and $\bar{t}$ disabled. This sample was used,
along with the nominal sample, as a template in the extraction of spin
correlation.  Backgrounds with two charged leptons in the final state were
simulated using \POWHEG-Box ($tW$), \SHERPA\ ($Z$+jets, diboson) and \MGMCatNLO
for other associated top-quark and top-quark pair production.  Backgrounds also
arise from events containing one prompt lepton from the decay of a $W$ or $Z$
boson and either a non-prompt lepton or a particle misidentified as a lepton.
These ``fake leptons'' were estimated using MC simulations and the result was
verified using a same-charge lepton control region in the data; the MC
distributions were scaled up by a small amount as a consequence.

Two types of signal events are considered, depending on whether a full
reconstruction of the $t\bar{t}$ system is performed, denoted here as
\emph{inclusive} and \emph{reconstructed} selections. The inclusive selection
is used for the $\Delta\phi$ and $\Delta\eta$ differential cross-sections.  It
is defined by requiring exactly one electron and one muon of opposite electric
charge and at least two jets, at least one of which must be $b$-tagged. The
reconstructed selection is used for the measurement of $\Delta\phi$ as a
function of the \ttbar invariant mass.  It has a more stringent $b$-tagging
requirement of at least two $b$-tagged jets and also requires that at least one
solution was found for the reconstruction of the \ttbar system.  Using the
inclusive (reconstructed) selection, 93\% (96\%) of the 177k (76k) selected
events are expected to be \ttbar events.  The data and prediction agree within
uncertainties for all kinematic observables studied.

In order to measure spin correlations as a function of the \ttbar invariant
mass at detector level, the kinematic properties of the event are reconstructed
from the identified leptons, jets, and missing transverse momentum. The top
quark, top antiquark, and reconstructed \ttbar system are built using the
neutrino weighting method.  The efficiency for \ttbar reconstruction is $80\%$. 

Events are corrected for detector effects using two definitions of particles in
the generator-level record of the simulation: parton level and particle level.
Parton-level objects are taken from the MC simulation history. Top quarks are
taken after radiation but before decay whereas leptons are taken before
radiation.  The measurement corrected to parton level is extrapolated to the
full phase-space, where all generated dilepton events are considered.  Fiducial
requirements are not made on the partonic objects so that the results at parton
level can be more easily compared to fixed-order predictions.  Particle-level
objects are constructed using a procedure intended to correspond as closely as
possible to the detector-level object and event selection.  Events are selected
at the particle level in a fiducial phase-space region with similar
requirements to the phase-space region in the detector. 

\newpage

Background-subtracted data are corrected for detector resolution and acceptance
effects using an iterative Bayesian unfolding procedure in order to create
distributions at particle (parton) level in a fiducial (full) phase-space.  The
binning for each observable is chosen in order to minimise the effect of
statistical fluctuations in the data as well as in the alternative \ttbar
samples which are used in the systematic prescription, as well as to account
for the experimental resolution.  The $\Delta\phi$ distributions are split into
four mass regions: $m_{\ttbar}<450$~\GeV; $450\leq m_{\ttbar}<550$~\GeV;
$550\leq m_{\ttbar}<800$~\GeV; and $m_{\ttbar}\geq800$~\GeV. 

The level of spin correlation observed in data is assessed by quantifying it in
relation to the amount of correlation expected in the SM.  This fraction of
SM-like spin correlation ($f_{\textrm{SM}}$) is extracted using hypothesis
templates that are fit to the parton-level, unfolded normalised cross-sections
from data. Two hypotheses are used: dileptonic \ttbar events with SM spin
correlation (the nominal \ttbar sample) and dileptonic events where the effect
of spin correlation has been removed (the nominal \ttbar sample where the top
quarks are decayed using \MADSPIN\ with spin correlations disabled).  The
extraction of $f_{\textrm{SM}}$ is performed in five observables: the inclusive
$\Delta\phi$ and $\Delta\phi$ in each of the four regions of $m_{\ttbar}$. The
total number of bins used in the extraction, $N$, depends upon the region of
$m_{\ttbar}$.  Systematic uncertainties on $f_{\textrm{SM}}$ are determined
using MC samples with different sources of systematic uncertainty, and the
unfolded spectra are used as pseudo-data.  The templates are fit to this
pseudo-data and the difference between the systematic $f_{\textrm{SM}}$ and the
nominal (i.e.  $f_{\textrm{SM}}=1$) is taken as the systematic uncertainty on
$f_{\textrm{SM}}$ due to that source.  The largest sources of systematic
uncertainty arise due to the modelling of the \ttbar process. 

For the inclusive result, the spin correlation extracted from the unfolded data
is significantly higher than the SM expectation at a significance of
$3.8\,\sigma$ without including theoretical uncertainties on the hypothesis
templates, and at $3.2\,\sigma$ when including these uncertainties.  The
central $f_{\textrm{SM}}$ value as a function of $m_{\ttbar}$ is found to
increase as a function of $m_{\ttbar}$, however, the uncertainties on
$f_{\textrm{SM}}$ are much larger than in the inclusive case and none of the
results deviate significantly from the SM expectation. 
 
A number of cross-checks were performed to attempt to explain the results in
terms of either the limitations of modelling of the \ttbar system or by
experimental effects not covered by the systematic uncertainties.  The
generators used in this analysis do not fully include NLO effects in the decays
of the top quarks, nor do they directly consider the effects of interference
between the initial and final states, but these effects are found not to be
relevant for the measurement and do not explain the observed deviation.  The
effect of the narrow-width approximation (NWA) on the $\Delta\phi$ observable
was investigated in an inclusive \ttbar + $tW$ dilepton phase-space using the
\textsc{Powheg-Box-Res} bb4l process and compared to the nominal \ttbar + $tW$
set-up and no significant differences were observed. It is therefore assumed
that, in the \ttbar phase-space of this measurement, the NWA in the templates
is not a limiting factor and does not explain the observed deviation.
Alternative templates for $f_{\textrm{SM}}$ extraction may be constructed from
samples used to evaluate systematic uncertainties.  In each case, the no-spin
template is derived by scaling the prediction of the alternative model (with
spin included) by the ratio of the no-spin and spin templates in the nominal
setup.  With the exception of the highest $m_{\ttbar}$ bin, which has large
statistical and systematic uncertainties, the $f_{\textrm{SM}}$ values remain
above 1 for all alternative templates.

The effect of next-to-next-to-leading order (NNLO) corrections in the
production was investigated by reweighting the $p_{\textrm{T}}(t)$ spectra in
\POWHEG + \PYTHIA 8 to NNLO fixed-order predictions and to observed
detector-corrected data spectra. The effect reduced the observed deviation
somewhat but was consistent with the scale uncertainties that are already
considered in the uncertainties on the hypothesis templates. Fixed-order NNLO
predictions recently became available for the observables under
study~\cite{Behring:2019iiv}.  The results of these predictions are closer to
the data than the NLO predictions, but still do not fully describe the observed
discrepancy.  An alternative differential prediction using a fixed
renormalisation and factorisation scale choice of the top mass and performing
an expansion in both QCD and electroweak (EW) couplings at NLO, made
specifically for these observables~\cite{Bernreuther:2015yna}, is used as a
template. This prediction agrees better with the data but has significant scale
uncertainties, leading to an $f_{\textrm{SM}}=1.03\pm0.13$, and is consistent
both with the result from using the \POWHEG + \PYTHIA 8 templates and with the
SM expectation.

Data is compared with various SM predictions in Figure~\ref{fig:spincorres}.
The disagreement between the data and the NLO predictions from MCFM and \POWHEG
+ \PYTHIA 8 can be clearly observed. The NNLO fixed-order prediction agrees
better with the data but still differs significantly. Finally, the expanded NLO
QCD + EW prediction agrees with the data within its large scale uncertainties.

\begin{figure}[htbp]
\centering
\hfil
\subfloat[]{\includegraphics[height=0.3\textheight]{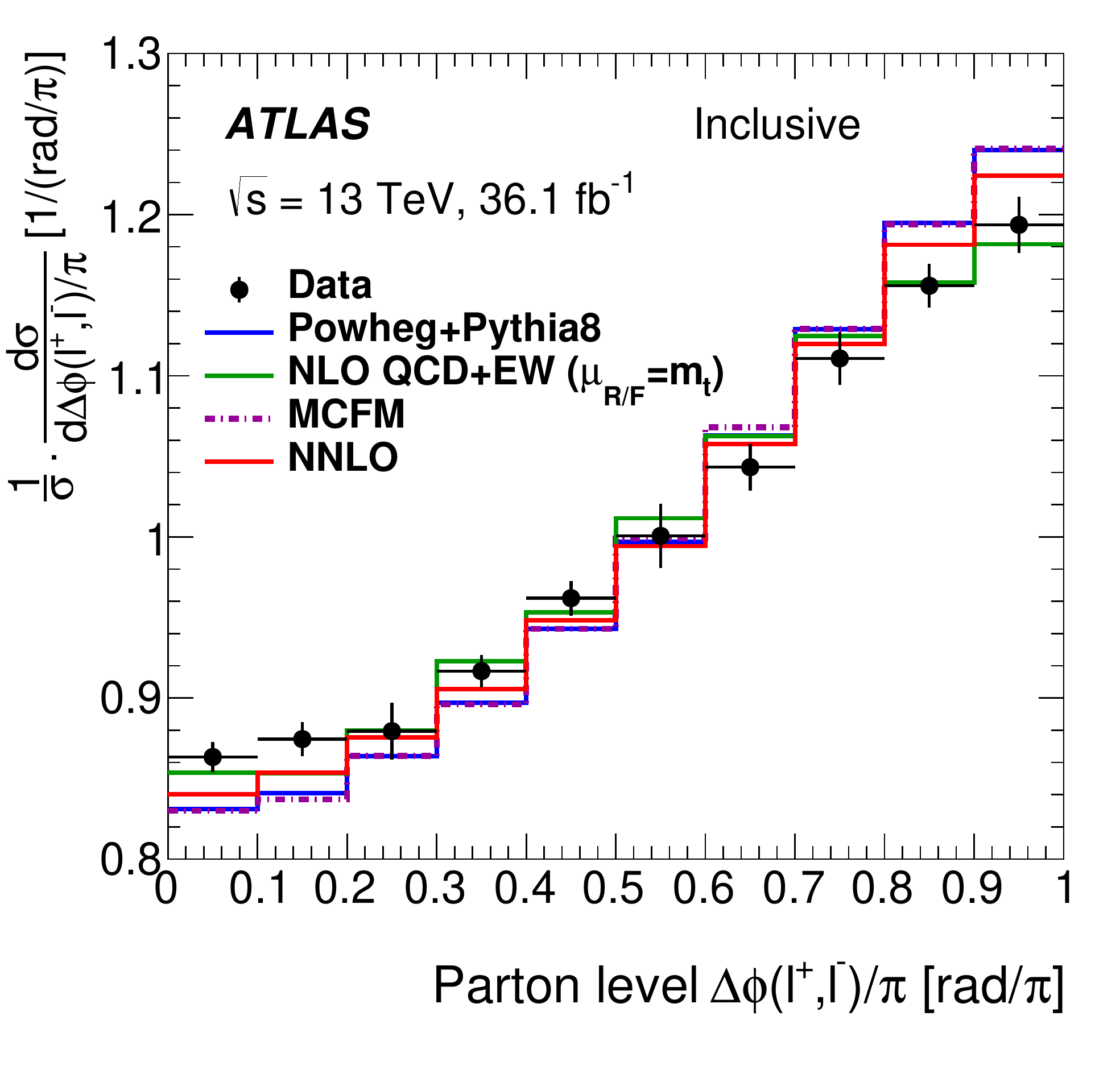}}
\hfill\hfill
\subfloat[]{\includegraphics[height=0.3\textheight]{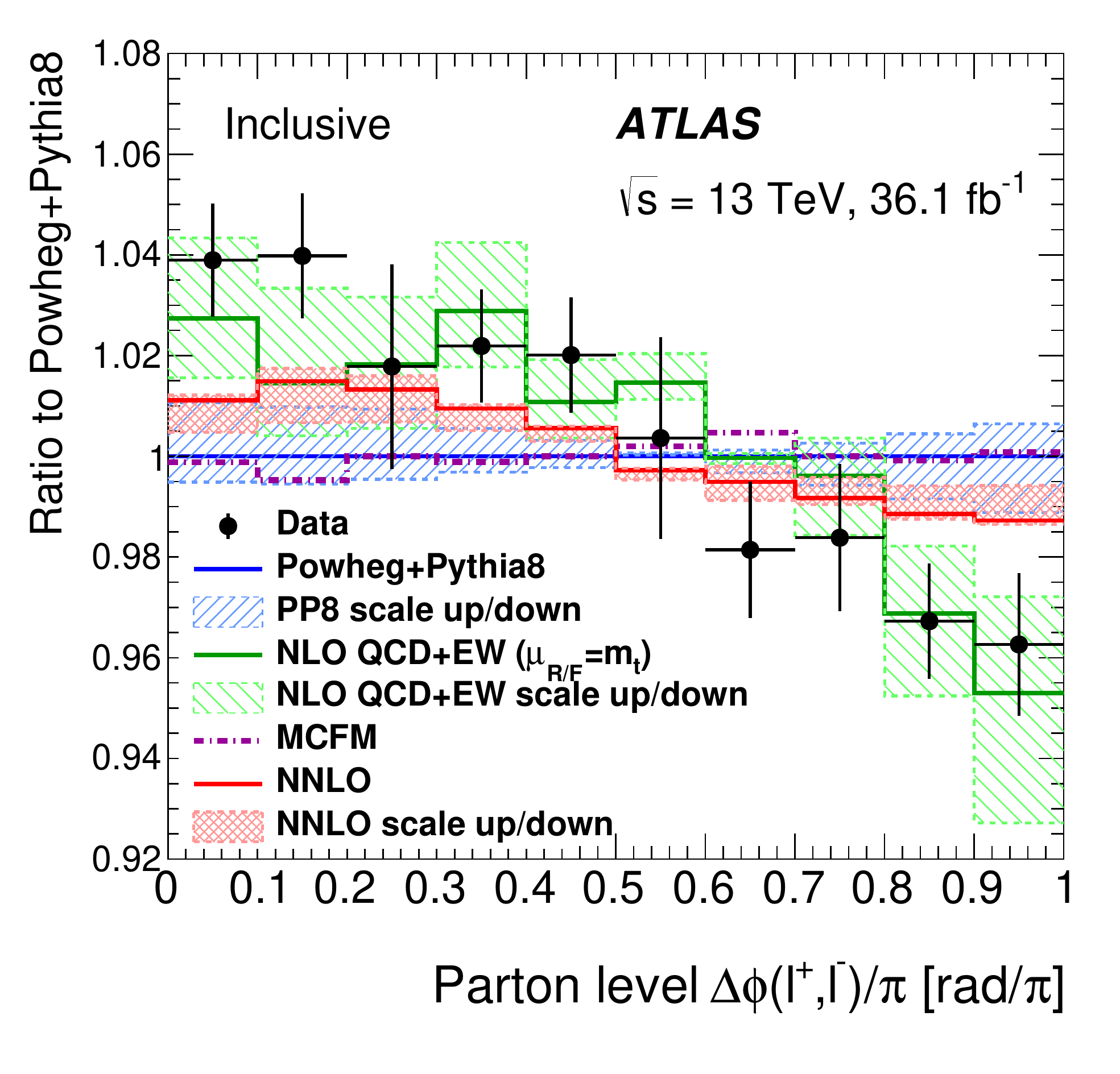}}
\hfill
\caption{%
Comparison of the unfolded $\Delta \phi$ distribution with theoretical
predictions for the inclusive selection; (a) normalized cross-section, (b)
ratio as compared with \POWHEG + \PYTHIA 8~\cite{spin}.}
\label{fig:spincorres}
\end{figure}
\FloatBarrier

\section{Status of other top-quark property measurements and summary}

The most precise top-quark related properties measured by the ATLAS
collaboration per LHC run energy are summarised and compared to the
corresponding theoretical expectations in Figure~\ref{fig:properties}.  The
properties are categorised as intrinsic properties of the top quark or as
properties of its production or decay.  ATLAS top-quark measurements can be
conveniently accessed through the TopPublicResults Wiki
page,\footnote{https://twiki.cern.ch/twiki/bin/view/AtlasPublic/TopPublicResults}
while the used theoretical predictions are further detailed in
Reference~\cite{properties}.

\begin{figure}[htbp]
\centering
\includegraphics[height=0.35\textheight]{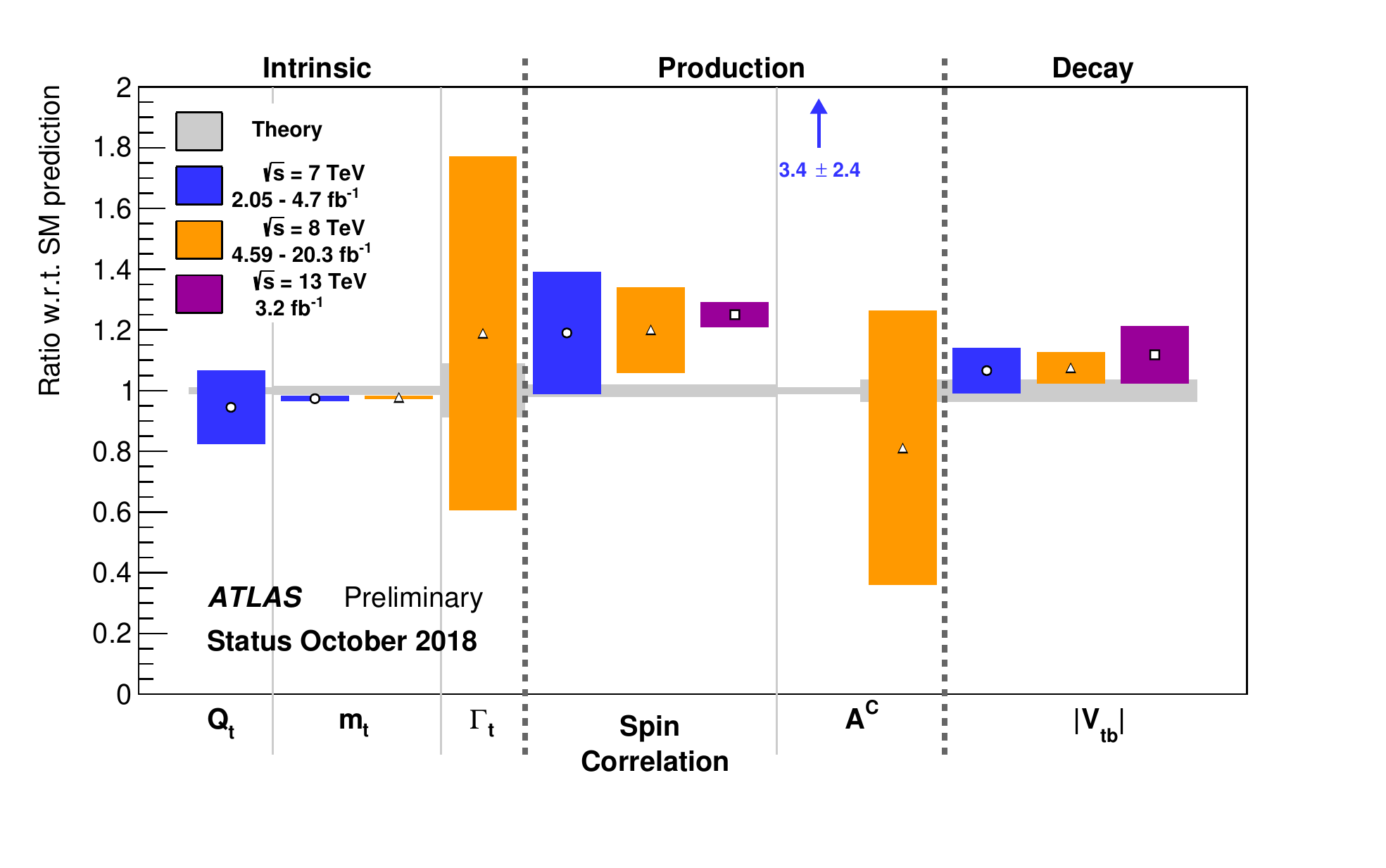}
\caption{Overview of top-quark properties measurement by the ATLAS
collaboration~\cite{properties}.} 
\label{fig:properties}
\end{figure}

The two most recent top-quark properties measurements released by the ATLAS
collaboration have been presented.  The normalised differential cross section,
\Robs, of top-quark pair production in association with an energetic jet is
presented as a function of the inverse of the invariant mass of the
\ttbarjet~system $\Rhos = 2 m_0/m_{\ttbarjet}$. The measurement is performed
using $pp$ collision data at a centre-of-mass energy of 8~\TeV{} collected by
the ATLAS experiment at the LHC in 2012.  The data sample corresponds to an
integrated luminosity of 20.2~$\mathrm{fb}^{-1}$.  The distribution of \Robs
observed in the semileptonic final state is unfolded to the parton and particle
levels.  The result from data unfolded to parton level is compared with the NLO
QCD predictions in two different renormalisation schemes.  The top-quark mass
extracted in the pole-mass scheme yields $\Mpole = 171.1 \pm 0.4 \stat \pm 0.9
\syst ^{+0.7}_{-0.3} \theo \GeV$.  The result for the running mass in the
\MSbar scheme is $m_t(m_t) = 162.9 \pm 0.5 \stat \pm 1.0 \syst ^{+2.1}_{-1.2}
\theo \GeV$.  The result for \Mrun suffers from a larger theoretical
uncertainty as compared with the pole mass.  This is due to a larger dependence
on the renormalisation and factorisation scales of the \MSbar scheme in the
most sensitive region close to the $\ttbarjet$ threshold.

Absolute and normalised differential \ttbar cross-sections have been measured
as a function of $\Delta\phi$ and $\Delta\eta$, between the two charged leptons
in the $e\mu$ decay channel using 13~\TeV\ data collected in 2015 and 2016. The
$\Delta\phi$ differential cross-section is also measured as a function of the
\ttbar invariant mass. None of the studied generators are able to reproduce the
normalised $\Delta\phi$ distribution.  A comparison was made with fixed-order
predictions at NNLO in QCD and in an expansion at NLO in QCD and EW couplings
with a fixed scale choice, with the former slightly improving the description
of the data, and the latter describing the data but with large scale
uncertainties.  An extraction of spin correlation was performed using the
normalised parton-level $\Delta\phi$ observable. The spin correlation was found
to be higher than that predicted by the SM as implemented in NLO MC generators
with a significance of $3.2\,\sigma$. However, the measurement agrees well with
the prediction by the expansion at NLO in QCD and EW couplings. 
 
\section*{Acknowledgements}

The author would like to thank Ren\'e Poncelet, Markus Schulze and Miriam Watson for 
useful discussions.
This work was partially funded by the European Research Council under the
European Union's Seventh Framework Programme ERC Consolidator Grant Agreement
n.~617185 (TopCoup) and
by the German Federal Ministry of Education and Research (BMBF) in FSP-103 under
grant n.~05H15PDCAA.


\begin{thebibliography}{99}
\bibitem{polemass} \coll{ATLAS}, \emph{Measurement of the top-quark mass in
$t\bar{t}+1$-jet events collected with the ATLAS detector in $pp$ collisions at
$\sqrt{s}= 8$ TeV}, \subm{JHEP}, \arxiv{1905.02302}{ex}.

\bibitem{mass} \coll{ATLAS}, \emph{Measurement of the top quark mass in the
$t\bar{t}\to$ lepton+jets channel from $\sqrt{s}=8$ TeV ATLAS data and
combination with previous results}, \pub{\EPJC 79 (2019)
290}{10.1140/epjc/s10052-019-6757-9}, \arxiv{1810.01772}{ex}.

\bibitem{properties} \coll{ATLAS}, \emph{Top Working Group Summary Plots --- Autumn 2018},
ATL-PHYS-PUB-2018-034, CERN, 2018, \pub{cds.cern.ch/record/2647993}{http://cdsweb.cern.ch/record/2647993}.

\bibitem{spin} \coll{ATLAS}, \emph{Measurements of top-quark pair spin
correlations in the $e\mu$ channel at $\sqrt{s} = 13$ TeV using $pp$ collisions
in the ATLAS detector}, \subm{\EPJC}, \arxiv{1903.07570}{ex}.

\bibitem{Behring:2019iiv} 
A.~Behring, M.~Czakon, A.~Mitov, A.~S.~Papanastasiou and R.~Poncelet,
\emph{Higher order corrections to spin correlations in top quark pair production at the LHC},
\arxiv{1901.05407}{ph}.

\bibitem{Bernreuther:2015yna} 
W.~Bernreuther, D.~Heisler and Z.~G.~Si,
\emph{A set of top quark spin correlation and polarization observables for the LHC: Standard Model predictions and new physics contributions},
\pub{JHEP 12 (2015) 026}{10.1007/JHEP12(2015)026}, \arxiv{1508.05271}{ph}.

\end{thebibliography}
\end{document}